# On the Limits of Design: What Are the Conceptual Constraints on Designing Artificial Intelligence for Social Good?

Jakob Mökander

**Abstract** Artificial intelligence (AI) can bring substantial benefits to society by helping to reduce costs, increase efficiency and enable new solutions to complex problems. Using Floridi's notion of how to design the "infosphere" as a starting point, in this chapter I consider the question: what are the limits of design, i.e. what are the conceptual constraints on designing AI for social good? The main argument of this chapter is that while design is a useful conceptual tool to shape technologies and societies, collective efforts towards designing future societies are constrained by both internal and external factors. Internal constraints on design are discussed by evoking Hardin's thought experiment regarding "the Tragedy of the Commons". Further, Hayek's classical distinction between "cosmos" and "taxis" is used to demarcate external constraints on design. Finally, five design principles are presented which are aimed at helping policy makers manage the internal and external constraints on design. A successful approach to designing future societies needs to account for the emergent properties of complex systems by allowing space for serendipity and socio-technological coevolution.

J. Mökander
Oxford Internet Institute, University of Oxford, Oxford, UK
e-mail: Jakob.mokander@oii.ox.ac.uk

## 1 Introduction

Artificial intelligence (AI) can bring substantial benefits to society by helping to reduce costs, increase efficiency and enable new solutions to complex problems (Taddeo and Floridi 2018). Because AI is malleable (Tavani 2002), the potential uses of computers are only limited by human creativity (Moor 1985). According to Floridi (2018), the main challenge in the age of AI is thus not innovation, but how to design the infosphere. This implies, Floridi suggests, that a normative cascade should be used to shape mature information societies that we can be proud of. To this aim, the "logic of design" puts the model (i.e. a blueprint), as opposed to the characteristics of the system, in the centre of the analytical process (Floridi 2017).

With Floridi's notion of design as a starting point, I propose to consider the question: what are the limits of design? The main argument of this chapter is that while design is a useful conceptual tool to shape technologies and societies, collective efforts towards designing future societies are constrained by both internal and external factors. The lack of goal convergence in distributed multi-agent systems makes it hard to define what the preferred blueprint or model should be in the first place (Helbing 2019). Moreover, the emergent properties of complex systems limit our ability to design them—even if a model is agreed upon (Hayek 1973a).

To support the main thesis, this chapter is divided into five sections. First, the Philosophy of Information is introduced to frame the subsequent discussion. Second, a literature review helps formulate a more precise research question: what are the conceptual constraints on designing AI for social good? Third, the internal constraints on design are discussed by evoking Hardin's (1968) thought experiment "the Tragedy of the Commons". Fourth, Hayek's (1973a) classical distinction between "cosmos" and "taxis" respectively is used to demarcate external constraints on design. Finally, five design principles aimed at helping policy makers manage the internal and external constraints on design are presented.

Although this chapter demonstrates the limits of design, it neither diminishes the importance nor the desirability of thinking about the future in terms of design. On the contrary, by highlighting the conceptual constraints on designing AI for social good, the aim is to contribute to the discourse about the dynamic relationship between innovation and governance. A successful approach to designing future societies, this chapter concludes, needs to account for the emergent properties of complex systems by allowing space for serendipity and socio-technological coevolution.

## 2 The Philosophy of Information and the Logic of Design

The world is experiencing a rapid process of digitalisation (Kolev 2018), through which ICTs have come to permeate all aspects of society from healthcare to dating (Cath et al. 2018a, b). Some societies not only use, but are dependent on ICTs for their very functioning (Floridi 2014). As part of this digital transformation, AI is increasing the potential impact of human actions to such an extent that our former ethical frameworks can no longer contain them (Jonas and Herr 1984).

As a result of what Floridi (2015) calls the blurring between reality and virtuality, human societies struggle to formulate positive visions or blueprints for the future. However, as Moor (1985) points out, policy vacuums are often products of conceptual vacuums. While philosophy is incapable of addressing factual issues, it can provide us with the necessary vocabulary to articulate problems and evaluate solutions (Bencivenga 2017). For the purpose of exploring the limits of design, I argue that the Philosophy of Information is a fruitful attempt to address this conceptual vacuum by adopting an information-centric level of abstraction.

An information-centric level of abstraction (LoAi) views the world in terms of the creation, storage, processing and manipulation of information as well as the relationships between informational entities (Floridi 2008). While a phenomenon can be described by many different LoAs, the appropriate LoA contains only the necessary conditions (Floridi and Sanders 2004). Because digital technologies provide new affordances for our creative design efforts (Floridi 2017), this chapter adopts LoAi to analyse the constraints on design.

By adopting LoAi, Information Ethics (IE) suggests that there is something more fundamental than life, namely being, and something more fundamental than suffering, namely entropy (Floridi 2014a, b). It follows that all entities have an intrinsic moral value and can count as moral patients (Hepburn 1984). The main takeaway, with regards to the limits of design, is that IE addresses agents not just as 'users' of the world, but also as 'producers' who are responsible for its well-being.

In *The Logic of Design as a Conceptual Logic of Information* (2017), Floridi argues that by shifting the focus from describing the system to designing the model, we can proactively participate in shaping the infosphere of tomorrow. Although the idea might seem radical, it is in fact already happening. Most sciences, including engineering and jurisprudence, do not only study their systems; they simultaneously build and modify them. According to IE, the possibility to design future societies also comes with a moral obligation to do so according to blueprints that allows for the flourishing of the entire infosphere. Next, I thus turn to explore the possibilities of designing AI for social good.

## 3   Artificial Intelligence and Design for Social Good

For the purpose of this chapter, AI is defined as "a resource of interactive, autonomous, and self-learning agency that can deal with tasks that would otherwise require human intelligence to be performed successfully" (Floridi and Cowls 2019). Using this definition, AI relates to design for social good in at least three different ways. First, AI can be the object of design, as developers attempt to design robust and trustworthy software (Russell et al. 2015). Second, AI can be the agent of design, since interactive, autonomous and adaptable systems influence and manipulate their environment (Floridi and Sanders 2004). Third, AI can be the mediator in the design process, serving as a tool to find more innovative and accurate solutions (Taddeo and Floridi 2018). Although this chapter focuses on the latter aspect of AI as a tool for design, all three cases will be considered in turn.

As an object of design, AI systems pose ethical risks related to bias, discrimination and transparency (Leslie 2019) as well as normative challenges like the transformative effects of recommender systems (Milano et al. 2019). While over 100 reports have proposed guidelines for ethical AI design, researchers have started to converge on a set of principles in line with classical bioethics, including beneficence, non-maleficence, autonomy, justice and explicability (Floridi and Cowls 2019). A discussion of which ethical principles to embed in AI is beyond the scope of this chapter. Instead, it suffices to acknowledge that some principles, like for example accuracy and privacy, conflict and require trade-offs for which there are no easy solutions (AI HLEG 2019). It is therefore essential to note that vague concepts like fairness and justice mask underlying political tensions (Mittelstadt 2019). These inherent conflicts create internal constraints on design. Or, as Russell et al. (2015) put it; in order to build robustly benevolent systems, we need to define good behaviour in each domain.

As an agent of design, AI systems challenge the view of moral agents as necessarily human in nature (Floridi and Sanders 2004). By adopting LoAi, however, IE does not discriminate between human and artificial agents. This has two important implications. First, expanding the class of moral entities to include artificial agents (as well as organisations and legal persons) allows for moral responsibility to be distributed (Floridi 2014). This is essential to ensure accountability in case some harm is caused by a distributed system, a scenario which is becoming increasingly plausible in information societies. Second, by disregarding both agents and their actions, IE shifts focus to the moral patient and the features of the infosphere that we want to see pursued or avoided (Floridi 2016a). Humans alone have, however, had difficulty converging on visions for a good AI society (Cath et al. 2018a, b). It is therefore hard to see how this becomes any easier as the class of moral agents is expanded.

Finally, AI can be a mediator in the design process insofar as it is applied as a tool to solve societal challenges (Cowls et al. 2019). In fact, AI can support the achievement of all UN Sustainable Development Goals (SDGs) (Vinuesa et al. 2019). Already, AI supports renewable energy systems and smart grids which are needed to address climate change. Moreover, AI can improve government in at least three ways: by personalising public services, making more accurate forecasts, and simulating complex systems (Margetts and Dorobantu 2019). However, many AI-based projects fail due to heedless deployment or poor design (Cath et al. 2018a, b). Some AI systems even end up having negative consequences, including the loss of employment opportunities following automation, or the increased inequalities resulting from the data economy (Vinuesa et al. 2019).

Since digital computers may theoretically carry out any operation which could be done by a human (Turing 1950), the number of potential use cases is infinite. However, the range of computable problems is not the same as the range of human problems (Weizenbaum 1984). It would therefore be pointless to extend the list of positive and negative examples of AI-applications. Instead, the lessons learned from AI as a tool for designing "good societies" indicate that policies concerning AI are subject to the same constraints as all governance. These include, but are not limited to, internal incoherence of preferences, inability to access resources, imperfect information and asymmetries between top-down directives and bottom-up incentives (Weiss 2011). Next, I therefore proceed to discuss two classical governance dilemmas, the tragedy of the commons and the distributed nature of knowledge in society, from a design perspective.

## 4 Collective Action Problems and the Internal Constraints on Design

In this section, I explore the internal constraints on design by, in line with IE, assuming that some states of the infosphere are morally better than others (Floridi 2014). This assumption is not controversial in itself, but deeply rooted in both consequentialist and deontological ethical traditions (Benn 1998). However, while such an assumption is a precondition for designing AI for social good, it does not address the hard question of which states are to be preferred.

On the one hand, the moral laws of IE enable an axiomatic evaluation from an ontocentric perspective (Floridi 2014). On the other hand, ethical principles vary depending on cultural contexts and the domain of analysis (Taddeo and Floridi 2018). Consequently, there is a tension between the ontocentric and the anthropocentric approach, since the latter is based on the principle that "good" and "evil" are not only identified by human beings but depend on human interests and perspectives. To act as stewards of the infosphere, humans would therefore need to overcome both our anthropocentric biases and manage our internal collective action problems.

Since the earliest days of human existence, communities have been faced with collective action problems. Popularised by Mancur Olson's seminal work *The Logic of Collective Action* (Olson 1965), the term describes situations in which all agents would be better off cooperating, but fail to do so because of individual incentives that discourage joint action. One example, provided by David Hume in *A Treatise of Human Nature* (Hume 1739) chronicles a village attempting to drain a meadow. It would be difficult, Hume assesses, for the villagers to concert and execute such a complicated task while each seeks a pretext to free himself of the trouble and lay the whole burden on others. It may hence seem that ruin is the destination toward which all men rush, each pursuing his own best interest.

In today's increasingly interconnected world, global challenges like climate change and nuclear proliferation are examples of collective action problems (Shackelford 2016) that threaten the future flourishing of both humanity and the larger infosphere. These challenges are global in a twofold sense; they concern humankind as a whole, and they can only be solved by humankind as a whole (Hofkirchner 2010). Consequently, any attempt to shape the future must address the questions: what should be maximized? and, how are collective action problems to be overcome?

The tragedy of the commons (TC) is an example of a collective action problem that is particularly applicable to shared environments. In *The Tragedy of the Commons* (1968), Hardin shows that it is "rational" for people to pollute the air, or pick flowers in national parks, as long as there is no personal cost involved. Moreover, neither appealing to conscientiousness nor introducing legislation seem to alleviate TC, since such measures either go counter to evolution (i.e. are not favoured by natural or sexual selection) or are hard to enforce. Hardin's insights about TC also translate into the digital realm. Artificial agents may, for example, exploit or pollute the infosphere (Greco and Floridi 2004). One example of TC in the infosphere is the excessive use of bandwidth by individuals or organisations, without any consideration of the needs of other users.

At the same time, AI and digital platforms provide agents with new tools to coordinate collective action (Helbing 2019). Digital technologies and infrastructure thereby contribute to a beneficial development in at least three ways. First, it is possible to shape an infrastructure that, although not morally good or evil in itself, can facilitate or hinder actions that lead to good or evil states of the system (Floridi 2016a). This could, for example, entail designing technologies in line with "infra-ethical values" like transparency and traceability to increase trust levels in society. Second, education can counteract the natural tendency to do the wrong thing (Hardin 1968). Since AI can both increase the quality of education and make it increasingly accessible and affordable (Meredith et al., p. 20), design of AI-based applications can strengthen our capacity to manage TC. Finally, digital entities are typically non-exhaustive and non-rivalrous (Yanisky-Ravid and Hallisey 2018). If, for example, a file is downloaded from the internet, this does not imply the destruction of the file itself (Greco and Floridi 2004). Thus, AI affords new solution-spaces in which to design for social good.

Although there is a strong case for shaping both AI and future societies, design efforts are constrained by lack of internal coherence. This lack of internal goal convergence is partly rooted in the conflicting impulses harboured by what Deleuze (1992) calls 'dividual' agents. Deleuze's "dividual" echoes Dawkins' (1976) claim that a human is not a singular coherent unit but rather a collection of selfish genes, each with its own motivations for survival. Although the theories of both Deleuze and Dawkins remain controversial, viewing society as a dynamic system consisting of "dividual" agents helps explain why it is subject to complex, conflicting and time dependent expressions of interests.

In conclusion, the above discussion has shown that the fallible ability of human agents to coordinate and achieve mutually beneficial outcomes limits our ability to manage global collective action problems (Lamb 2018). In *Artificial Intelligence and the 'Good Society'* (2018), Cath et al. highlight this constraint by concluding that although a proactive design of AI-policies is possible, an overarching political vision for what a "good AI Society" should look like is lacking. Given that this shortcoming, as demonstrated by Hardin, is deeply rooted in human nature, the difficulty to collectively construct, agree upon and implement preferable "models" remains an internal constraint to design AI for social good. This does not mean that TC cannot be managed. In fact, scholars like Nobel laureate Elinor Ostrom have proposed institutional mechanisms to overcome TC. Such strategies and mechanisms will be examined in section five. Next, however, I now turn to explore external limits of design.

## 5    Cosmos, Taxis and the External Constraints on Design

Floridi (2009) claims that the best way of tackling the new ethical challenges posed by ICTs is from a more inclusive approach, i.e. one that does not privilege the natural. The use of the term 'natural', as opposed to created or artificial, presupposes the existence of an objective reality. Following a long philosophical tradition ranging from Aristotle and Confucius to Spinoza, IE could thus be viewed as a naturalist theory (Hongladarom 2008). On the one hand, the fact that something is considered natural does not imply that it is morally good (Benn 1998). On the other hand, the existence of an objective reality limits the prospects of designing the same reality. Human beings operate in a space of affordances and constraints provided by technological artifacts and the rest of nature. Humans are thus "free with elasticity" (Floridi 2015). Given that social hierarchies can be viewed as structures that emerge from the interactions of individual agents (Collins 1994), similar constraints on design apply to societies. The question is therefore what is to be considered "natural" in the infosphere?

To address the above-mentioned question, it is helpful to recall Hayek's classical distinction between "cosmos" and "taxis". In *Law, Legislation and Liberty* (1973), Hayek first defines "order" as "elements of various kinds being so related to each other that we may learn, from our acquaintance with some spatial or temporal parts of the whole, to form correct expectations concerning the rest". Put differently, there must be some order and consistency in life. Subsequently, Hayek distinguishes between two kinds of order. Cosmos represents natural, spontaneous orders, whereas taxis represent manmade, constructed or artificial orders. Finally, Hayek suggests that cosmos and taxis follow different logics. Cosmos, or the natural, is complex, emergent, and cannot be said to have a purpose. "Emergence", in this case, implies that an entity can have properties its parts do not have on their own, and that randomness can give rise to orderly structures (Corning 2010). This leads Hayek to conclude that complex systems, including emergent phenomena like biological organisms or human societies, are not only hard to understand, but also hard to control and govern.

Hayek's insights carry significant relevance to the prospect of designing AI for social good. First, it is difficult to predict what outcomes even benign attempts to nudge society in a certain direction will have, given that complex systems often are subject to nonlinear feedback loops. Naive claims to "program good ethics into AI" (see e.g. Davis 2015) can thus be dismissed as too simplistic. Second, it is difficult to govern a multi-agent system, given that knowledge is distributed in society (Hayek 1945). The designer therefore depends on that other individuals, who are to cooperate, make use of knowledge that is not available to the central authority. Even if principles for how to design AI for social good can be established, questions about *how* these factors should be evaluated and by *whom* remain unanswered (Cowls et al. 2019). Third, the more complex a system is, the more the design effort is subject to unknown circumstances. Thus, while ICTs provide new tools for collecting, analysing and manipulating information (Taddeo and Floridi 2016), the emerging infosphere will be even more difficult to design than previous socio-technical environments.

Hayek's hesitation towards the ability of humans to control and design future societies is healthy. In fact, Alan Turing (1950) pointed out that it appears impossible to provide rules of conduct to cover every eventuality, even those arising from bounded systems like traffic lights. Since Turing's seminal paper, the positivistic reliance on concepts like "organisation" and "rationalism" has been further deconstructed by French postmodernists like Derrida and Foucault (Peet and Hartwick 2015). This, however, does not mean that we should accept a "laissez-faire" view on public policy. While foresight cannot map the entire spectrum of unintended consequences of AI systems, we may still identify preferable alternatives and risks mitigating strategies (Taddeo and Floridi 2018). We must therefore avoid the dual trap of surrendering to either technological determinism or moral relativism.

Given that societies are increasingly delegating risk-intensive processes to AI systems, such as granting parole, diagnosing patients and managing financial transactions (Cath et al. 2018b), we cannot abstain from engaging in design. Rather, design needs to be understood as a dynamic process of shaping technologies, policies and spaces in a complex, distributed and emergent multi-agent systems which are subject to constant feedback loops that alter the relationships between the entities. It is therefore essential that designs include an architecture for serendipity and leave room for continuous coevolution between and within socio-technical systems (Reviglio 2019).

Serendipity is understood as the art of discovering things by observing and learning from unexpected encounters and new information (Reviglio 2019). Since serendipity favours pluralism and innovation, while disfavouring efficiency and security, it is a design concept that resonates with the motto of IE "let a thousand flowers blossom". An architecture that allows for serendipity is also consistent with the nature of Hayek's cosmos. Consequently, although spontaneous orders impose external constraints on design, acknowledging and incorporating the need to balance factors like personalisation, generalisation and randomisation can help facilitate the design of AI for social good. Thus, the main point is that efforts to design future societies can be more successful by accounting for both the laws of cosmic orders and the complex characteristics of emergent phenomena.

# 6 The Pilgrim's Progress

The fact that a system is flawed does not imply that any other configuration of the system would be better. At the same time, the mere existence of a system does not imply that it cannot be improved. Policy makers are therefore continuously striking a balance between respecting the complexity of emergent systems, on the one hand, with efforts to design deliberate orders to further specific goals, on the other hand. While this chapter has identified both internal and external constraints on human efforts to design, it also acknowledges that efforts to design AI for social good can help shape future societies and that such efforts can be implemented more or less successfully. How then can policy makers account for the limits of design while designing purposeful policies? Although answering this question is beyond the scope of this chapter, at least five design principles follow directly from the above analysis of the limits of design. These are presented below to serve as a starting point for guiding policy makers in their efforts to design AI for social good.

## *6.1 Holistic Design*

Complex problems cannot be solved by breaking them up into solvable pieces because the different parts tend to interact in non-linear ways (Lamb 2018). While an analytical approach attempts to break down a problem and address its individual components, a holistic approach attempts problem solving at the same level of complexity as the issue itself (Holland 2014). Taking a holistic approach to design is particularly important when adopting LoAi, both because the class of moral agents is expanded and because the concept of "environmental flourishing" becomes increasingly abstract. A holistic approach is synonymous with what Meadows and-Wright (2009) calls "system thinking", i.e. that compromises and trade-offs are necessary. Moral evil is therefore, as Floridi (2010) puts it, unavoidable and the real effort lies in limiting it and counterbalancing it with more moral good.

## *6.2 Dual System Approach*

Any design solution proposed for a particular problem has to fight its way through two complex systems: first the problem-solving system, then the problem system (Lamb 2018). If, for example, a company wants to boost growth, the management must not only calibrate its business processes to beat the competition, but also convince its own organisation about the proposed changes. Similarly, policy makers need to anchor policies and blueprints for future societies in consultation with academic experts, the private sector and the civil society early and consistently. This entails that those affected by the rules can participate in modifying them (Ostrom 1990). A necessary, but insufficient, criteria for successful design is thus that blueprints are shared and supported by different agents in distributed systems and coherent with the spontaneous orders that constrain the solution-space.

## 6.3    Gradual Implementation

It is impossible to completely replace the spontaneous order with established organisation (Hayek 1973b). Put differently, society rests and will continue to rest on both spontaneous orders and on deliberate organisation. By applying gradual implementation of design strategies, policy makers can monitor system feedback and adapt to it. Although conceptually based on the experimental method of scientific knowledge creation, gradual implementation does not presuppose the possibility to acquire perfect knowledge or absolute truth. On the contrary, there is no greater error in science than to believe that just because some mathematical calculation has been completed, some aspect of nature is certain (Whitehead 1929). Consequently, any design effort must balance visionary power with humility even when practicing gradual implementation.

## 6.4    Tolerant Design

There is a tension between the ontocentric moral laws of Information Ethics, on the one hand, and anthropocentric values of the human agents who consciously attempt to design the infosphere, on the other hand (Hofkirchner 2010). This dilemma of toleration is well known within the social sciences; how can an agent be respectful towards another agent's choices while attempting to influence or regulate the same choices from an altruistic perspective? (Floridi 2015). Efforts to design AI for social good must therefore abide by what Floridi (2016b) calls "tolerant paternalism" or "responsible stewardship". In practice, this means that policy makers can influence individual agents positively through "nudging" on an informational level, while safeguarding toleration and respect to individual preferences on a structural level.

## 6.5    Design for Serendipity

There is a tension between the accuracy of a system and the extent to which it allows for serendipity, a key feature of environmental flourishing (Reviglio 2019). In essence, designing information architectures for serendipity increases the diversity of entities and encounters. Serendipity can thus be conceived as a design principle able to strengthen pluralism (Reviglio 2019). Pluralism is particularly important in self-organised systems, i.e. phenomena in which the order is not the result of an external intervention, but the outcome of local mechanisms iterated along thousands of interactions (Caldarelli and Catanzaro 2012). Given that the moral laws of IE do not discriminate between agents of different nature, increased plurality in biological, informational and hybrid systems contribute to the flourishing of the infosphere. Consequently, good design architecture must leave room for serendipity by balancing personalisation, generalisation and randomisation.

# 7 Conclusions

The digital revolution has led to a reontologisation of the world (Floridi 2011). In particular, the merger of the virtual and physical realities has led to a decoupling of hitherto coupled concepts, like use and ownership or location and presence (Floridi 2017). As humans become increasingly dependent on ICTs, we simultaneously contribute to shaping the infosphere. Not only engineers and software developers, but also users of technologies and citizens in general, help define the human "onlife" experience through their actions, choices and designs. Consequently, we inevitably shape both the technological systems we use and the social structures in which we live. The question, therefore, is whether we do so unknowingly or through conscious design?

At the same time, there are limits to design. First, design is constrained by conflicting internal interests. The logic of design presupposes a blueprint, or a model, which subsequently identifies a structure which belongs to a system under transformation (Floridi 2017). However, given that both "dividual" agents experience conflicting impulses (Deleuze 1992), and that societies are subject to collective action problems (Hardin 1968), a design model is difficult to agree upon. The lack of internal coherence thus constrains our attempts to design AI for social good from within. This internal limitation raises questions like: Which model should be used? How can tensions between partially conflicting normative values, like e.g. privacy and accuracy, be managed? What does a good AI society mean? Who is to decide? And, by what mechanism are design efforts coordinated?

Second, design is constrained by the spontaneous orders of the external environment. Any collaborative effort, including design, will thus always rest on both deliberate organisation and the abstract laws of complex systems (Hayek 1973a). Both the distributed nature of knowledge in society and the need for logic of lower-order systems to be compatible with higher-order systems put external constraints on our ability to design AI for social good. These external limitations raise questions like: Which are the domain-specific limitations imposed by cosmos on human design and organisation? How can systems be designed for accuracy and efficiency, while allowing for serendipity? And, how do we mitigate the risks associated with non-linear changes in complex systems when attempting to design interventions based on limited information?

While it is beyond the scope of this chapter to answer the questions posed above, some guidance for how policy makers can support human efforts to design AI for social good has been identified. First, design approaches need to be holistic and gradually implemented, both to anchor visions for the future amongst broader populations and to enable continuous evaluation of the non-linear effects of changes in complex socio-technical environments. Moreover, a tolerant paternalism must avoid optimising design efforts on individual system properties like efficiency, freedom or equality alone. Rather, design efforts need to continuously manage tensions between conflicting values through robust multiple-criteria analysis that regards the potential impact of unknown parameters. Strategies to achieve this aim could, for example, include allowing for generalisation alongside personalisation, or to introduce randomness in systems designed for accuracy.

The limitations identified in this chapter do not imply that attempts to design AI for social good are futile or undesirable. In fact, the opposite is true; properly designed, AI can contribute to the flourishing of the entire infosphere. It is, however, only by acknowledging the limits of design, and accounting for these constraints by adopting strategies that allow for plurality, serendipity and emergent phenomena, that mankind will be conceptually equipped to design AI for social good.